\documentclass[11pt]{article}

\usepackage{amsmath}
\usepackage{amssymb}
\usepackage{amsthm}
\usepackage{graphicx}
\usepackage[margin=1in]{geometry}
\usepackage{hyperref}

\newcommand{\wc}{\omega_c}

\def\be{\begin{equation}}
\def\ee{\end{equation}}
\def\bea{\begin{eqnarray}}
\def\eea{\end{eqnarray}}

\def\rme{\mathrm{e}}

\def\rme{\mathrm{e}}

\begin{document}
\title{\bf{The Thermodynamic Limit and the Landau degeneracy}}

\date{\today}

\author{St\'ephane Ouvry$^{*}$  }

\maketitle

\begin{abstract}

Usually, thermodynamic quantities are calculated by first regularizing the system in a confining box, which may have hard walls or periodic boundaries. One then studies the finite system and takes the thermodynamic limit by letting the box grow indefinitely.
Here, we discuss a regularization scheme devised in the late 1980s [1] to study the thermodynamic equations of an anyon gas [2]. Instead of confining the system with walls, the method adds a smooth harmonic potential. Quantities are evaluated in its presence, and the potential is then gradually removed. A suitable prescription for taking this limit is needed to obtain meaningful thermodynamic results.
\end{abstract}

\section*{Introduction}

We address the question of taking the thermodynamic limit by  confining a many body quantum system in a harmonic well and then letting the strength of the well go to zero.
First, as an illustration, we show how to recover the right scaling for the spectral degeneracy of a 2-dimensional particle coupled to a perpendicular magnetic field, the so called Landau degeneracy, without relying on semi-classical arguments. We then generalize these considerations to Riemann Zeta functions regularizations and their interpretation in quantum mechanics. 

\section*{1) The thermodynamic limit prescription}

Consider in full generality a quantum 1-body spectrum $\epsilon_i$, $i=1,\dots,k$ and the associated  1-body partition function
\[
Z_{\beta}=\sum_{i=1}^{k} \rme^{-\beta\epsilon_i}
\]
The thermodynamical potential (i.e. the logarithm  of the grand partition function which has to scale in the thermodynamic limit as the volume of the system)  for a gas of bosons in such a 1-body spectrum  is  
\[
\log\prod_{i=1}^{k}\frac{1}{1-z\,\rme^{-\beta\epsilon_i}}=\sum_{n=1}^{\infty} c_n\, z^n,
\]where the cluster expansion coefficients $c_n$ have the simple form \[
c_n=\frac{Z_{n\beta}}{n}.
\]
This is   the case for many-body spectra obtained  by  filling independently  1-body eigenstates   with occupation numbers $n_i$, in  the particular   boson case $n_i=0,\ldots,\infty$ (in the Fermi case $n_i=0,1$ there would be an additional $(-1)^{n-1}$ factor).
\subsection*{Thermodynamic limit prescription}
The thermodynamic limit prescription in $d$ dimensions arises from comparing the thermodynamical potential of a free gas of bosons\footnote{Note that all which is said here is equally valid for a gas of fermions  and leads to the same prescription.} confined in a $d$ dimensional  harmonic well  where $\omega$  the strength of the well is  understood to be small $\omega\to 0$ to the thermodynamical potential  of the same free gas of bosons now confined in a box of size $L^d$ where $L$  a 1d length is understood to be  large  $L\to\infty$.

On the one hand for a gas of bosons in a $d$ dimensional harmonic well  one has
\[
Z_{\beta}(\omega)=\left(\frac{\rme^{\beta\omega}}{\rme^{\beta\omega}-1}\right)^{d}\Rightarrow
c_n(\omega)=\frac{Z_{n\beta}(\omega)}{n}.
\]

On the other hand for a gas of bosons in a $d$ dimensional  box
with in the large $L\to\infty$ limit a 1-body density of states 
\begin{equation}\label{11}
\rho_d(\epsilon)=\frac{1}{\Gamma(d/2)}\,(\frac{L}{\sqrt{2\pi}})^{\,d}\,\epsilon^{\,d/2-1}
\end{equation}
where $L^d$  stands  for  the $d$-dimensional volume\footnote{$\sqrt{2\pi\beta}$ is the thermal wavelength.}  one has
\[
Z_{\beta}(L)=\int_0^\infty \rho_d(\epsilon)\,\rme^{-\beta\epsilon}\,d\epsilon
= \big(\frac{L}{\sqrt{2\pi\beta}}\big)^d
\Rightarrow  c_n(L)=\frac{Z_{n\beta}(L)}{n}
\]

In the thermodynamic limit  $\omega\to 0$ the harmonic well cluster coefficients become \[
\lim_{\omega\to 0}c_n(\omega)\simeq\frac{1}{n}\left(\frac{1}{n\beta\omega}\right)^{d}
\] As already said they should coincide with  the large box   cluster coefficients $c_n(L)$ therefore
\[
\frac{1}{n}\left(\frac{1}{n\beta\omega}\right)^{d}\ \longrightarrow \frac{1}{n} \big(\frac{L}{\sqrt{2\pi n\beta}}\big)^d
\]
It implies at order $n$ in the cluster expansion the \emph{\bf {thermodynamic  limit prescription}}
\begin{equation}
\boxed{\;\frac{1}{\sqrt{n}\beta\omega}\ \longrightarrow\ \frac{L}{\sqrt{2\pi\beta}}}\label{thecrux}
\end{equation}
which can  be used for computing  the thermodynamic equations of  interacting quantum many body systems in any dimension.  

The thermodynamic limit prescription (\ref{thecrux}) has been  extensively used in 2d in the context of the lowest Landau level (LLL) {\rm LLL}-anyon model \cite{moi} allowing for  a proper \rm LLL-anyonic gas thermodynamic     which does interpolate between the \rm LLL-Bose gas and \rm LLL-Fermi gas limits. It has also been used \cite{moiter} for the same purpose in 1d in the context of the Calogero model. Further one has been able to show  that in the vanishing magnetic field limit  the thermodynamic of the \rm LLL-anyon model is mapped on those of the Calogero model \cite{Ouvry1999} (for a review on all these efforts  see  \cite{moibis}). As a remark this mapping has also been shown to hold    at the level of  the exact $N$-body eigenstates \cite{OuvryPolychronakos2018}.

This  prescription has also  been used in other many body quantum  contexts \cite{Castin}, \cite{Drummond}.  It has also been  discussed on more formal grounds in \cite{Olaf}.

From now on we will concentrate on the  1-body case, i.e., the $n=1$ cluster coefficient prescription, to illustrate the usefulness of the mechanism at hand by showing how it  allows to retrieve in 2d the proper {\rm LLL} degeneracy in a purely quantum context  without relying on the usual semi-classical arguments. Also how  the priori awkward $B\to 0$ limit in the {\rm LLL} if properly taken allows for unveiling the above mentionned {\rm LLL}-anyon/Calogero thermodynamics mapping. We will then generalize these considerations to Zeta functions at negative entries and their possible interpretation  in a quantum mechanical context.

\section*{2) Application to the Landau degeneracy in the {\rm LLL}}

\noindent\textbf{1) Landau problem in 2-dimensional  plane:}

Consider a 2d charged particle $e$ in a perpendicular magnetic field $B$  with $eB\ge 0$ so that $ \omega_c={eB/ 2}$ is half the cyclotron   frequency (we work in the units $\hbar=m=1$).
The 1-body  Landau Hamiltonian 
\[
H=-{2}(\partial-{\omega_c\over 2}\bar z)(\bar\partial+{\omega_c\over 2}z) +\wc
\]
has  the {\rm LLL} eigenstates  
\[
\psi_l=z^{l}\rme^{-{1\over 2}\omega_c z\bar z}\quad l\ge 0
\]  
where $l\ge 0$ is the orbital momentum quantum number.

The 1-body {\rm LLL} energy \begin{equation} \omega_c\nonumber \end{equation}  is $l$-independent and therefore infinitely degenerate. One well-known claim is that the infinite  number of {\rm LLL} eigenstates, the so-called Landau degeneracy,   scales as  the infinite surface $S$  of the plane 
\[{B S\over\phi_0}\]
i.e.,  as the flux of $B$ through the plane counted in unit of  the flux quantum
 $\phi_0=2\pi/e$. We note that if in the {\rm LLL} there is only one quantum number $l$   the system is still bidimensional as encoded in the  surface scaling of the  degeneracy.

This claim  is traditionally supported by  semi-classical/hand-waving arguments: consider the radius\footnote{This radius  coincides with those of the classical cyclotron orbit obtained from Newton's equation $mv^2/r_l = evB$ and saying that  in the symmetric gauge where the  vector potential  $A_{\phi}=Br/2$ the canonical classical angular momentum $r_l (m v-e A_{\phi})$ is quantized to be  $l$.} $r_l=\sqrt{2l/(eB)}$ where the probability density $\psi_l\bar\psi_l$ is maximum, since  $\pi r_l^2 B/\phi_0 = l$, one can interpret $l$ as the flux of $B$ through the $l$-orbit of radius $r_l$ counted in unit of $\phi_0$. Then when $l\to\infty$ the total number of {\rm LLL}-eigenstates  acquires the meaning of being the flux of $B$ through the entire plane counted in unit of $\phi_0$.

Let us now show how this scaling can be retrieved  in a purely quantum mechanical context using solely the thermodynamical limit prescription discussed above. To achieve this goal one  focuses on the
 {\rm LLL} partition function
\begin{equation}
\label{LLL}
Z_{\rm  LLL}=\sum_{l=0}^{\infty}\rme^{-\beta\wc}
\end{equation}
What is the meaning of the infinite degenerate sum in (\ref{LLL})? Written as such it is an ambiguous, meaningless quantity. To remove this ambiguity as advocated above add a 2d harmonic well, compute the {\rm LLL} partition function in the presence of the harmonic well, then take the thermodynamic limit by letting the well vanish.
 
 In the presence of a harmonic well the {\rm LLL} spectrum becomes the harmonic {\rm LLL}  spectrum
\begin{equation}\nonumber
\omega_c\to \omega_t + l(\omega_t-\omega_c)
\end{equation} where $\omega_t =\sqrt{\omega^2+\omega_c^2}$.
The {\rm LLL} partition function becomes  the harmonic  LLL partition function
\begin{equation}\label{omega}
Z_{\rm LLL}\to Z_{{\rm LLL},\omega} = \sum_{l=0}^{\infty} \rme^{-\beta(\omega_t + l(\omega_t-\omega_c))}
= \frac{\rme^{-\beta\omega_t}}{1-\rme^{-\beta(\omega_t-\omega_c)}}
\end{equation}
Take now the thermodynamic  limit $\omega \to 0$ (i.e., strictly speaking $\beta\omega\to 0$)
\begin{equation}\label{limit}
\lim_{\omega\to 0} Z_{{\rm LLL},\omega}
\simeq \frac{\rme^{-\beta\omega_c}}{\beta(\omega_t-\omega_c)}
\simeq \frac{\rme^{-\beta\omega_c}}{\beta\omega^2/(2\omega_c)}
\end{equation}
In (\ref{limit}) one has to give a meaning to the diverging $1/\omega^2$ term. For this use  the thermodynamic  limit prescription (\ref{thecrux}) which says that when $\omega\to0$ the $n=1$ cluster coefficient becomes   \begin{equation}
\frac{1}{(\beta\omega)^2} \to \frac{L^2}{2\pi\beta}\label{prescrit}\end{equation} i.e., the 2d harmonic well partition function becomes the  2d free partition function. It follows  that
\begin{equation}
\lim_{\omega\to 0} Z_{{\rm LLL},\omega}
\simeq \beta(2\omega_c)\,\frac{L^2}{2\pi\beta}\, \rme^{-\beta\omega_c}
= eB\,\frac{L^2}{2\pi}\, \rme^{-\beta\omega_c}\nonumber
\end{equation}
If one  ignores the harmonic well one  ends up  with
\begin{equation}\label{LLLbis}
Z_{\rm LLL} = {B S\over\phi_0}\, \rme^{-\beta\omega_c}
\end{equation}
 and
for the associated density of states \begin{equation} \rho_{\rm LLL}(\epsilon)={B S\over\phi_0}\,\delta(\epsilon-\omega_c)\label{22}
\end{equation}
The Landau degeneracy $BS/\phi_0$ where the surface $S=L^2$ of the plane   indeed materialized, has been here rederived  by purely quantum arguments. What happened is that the degenerate sum in (\ref{LLL}) became tractable because a magnifying lens, i.e., the harmonic well, was applied, and when the thermodynamic limit was taken, thanks to the thermodynamic limit prescription,  the proper  expression was recovered with the right surface scaling.

\bigskip

\noindent\textbf{2) $B\to0$ limit:} 
\noindent If now one wants to  take the $B\to0$ limit\footnote{It is a bit awkward, we are in the LLL and then we take the $B\to0$ limit. It will become clear later why one does so.} (i.e., strictly speaking $\beta\omega_c\to 0$) in (\ref{LLLbis})  again we are  facing the same type of difficulties, namely  an
  ambiguous, meaningless expression: what means $BS/\phi_0$ when $B\to0$ while $S\simeq\infty$?
 Again apply the same kind of ideas as above:  remove the ambiguity by adding   a    2d harmonic well as a long distance regulator, i.e.,  instead of (\ref{LLL})  use rather (\ref{omega})  where the $B\to0$ limit can be done safely to obtain
\[
\lim_{B\to0} Z_{{\rm LLL},\omega} = \frac{\rme^{-\beta\omega}}{1-\rme^{-\beta\omega}}
\]
Then take the thermodynamic  limit $\omega\to0$
\[
\lim_{\omega\to0}\ \lim_{B\to0} Z_{{\rm LLL},\omega} \simeq \frac{1}{\beta\omega}
\]
and use the very same thermodynamic  limit prescription  (\ref{thecrux}) as already used in (\ref{prescrit}),  here $1/(\beta\omega)\to L/\sqrt{2\pi\beta}$. It follows that
\begin{equation}\nonumber
\lim_{\omega\to0}\ \lim_{B\to0} Z_{{\rm LLL},\omega}
\simeq \frac{L}{\sqrt{2\pi\beta}}
\end{equation} where 
${L}/\sqrt{2\pi\beta}$  is now  the 1d free partition function $Z_{\rm 1}$.
If one  ignores the harmonic well one ends up with
\begin{equation}\label{88}
\lim_{B\to0} Z_{\rm LLL} = Z_{\rm 1}\Leftrightarrow \lim_{B\to0}{B S\over\phi_0}\, \rme^{-\beta\omega_c}= \frac{L}{\sqrt{2\pi\beta}}
\end{equation}
or equivalently   with  the  {\rm LLL}-density of states  becoming  the free 1d density of states $\rho_1(\epsilon)$   in  (\ref{11})  \begin{equation}
\label{stunning}\lim_{B\to0} \rho_{\rm LLL}(\epsilon) =  \rho_1(\epsilon)\Leftrightarrow\lim_{B\to0} \frac{BS}{\phi_0}\delta(\epsilon-\omega_c) =  \frac{1}{\Gamma(1/2)}\,\frac{L}{\sqrt{2\pi}}\,\epsilon^{\,-1/2}\end{equation}
The equations (\ref{88}) and (\ref{stunning}) written as such seem a bit awkward. They become meaningful if one keeps in mind they have been obtained via the harmonic well regularization. It amounts to say that the degenerate spectrum of a 2d  particle in the {\rm LLL} of an external magnetic field becomes in the  limit where the field vanishes  the continuous spectrum of a  particle moving freely on a 1d line.  Indeed  one has followed the successive steps \[\omega_c\to_{\omega\ne 0}  \omega_t + l(\omega_t-\omega_c)\to_{B\to0}  (l+1)\omega\to_{\omega\to 0} k^2/2 \] where $k$ stands for the free momentum of a 1d particle on a line with free spectrum $k^2/2$. For $\omega$ small   the spectral gap  is $\omega^2/(2\omega_c)$ in the former $B\ne 0$ case whereas  it is $\omega$ in the latter $B\to 0$ case. As  far as the  partition functions are concerned,  moving from (\ref{LLL}) to (\ref{LLLbis}) then   from (\ref{LLLbis}) to (\ref{88}) used the same harmonic well  LLL partition function (\ref{omega}) and the same thermodynamic prescription (\ref{thecrux}) with the end results   differing because of  the different spectral gaps. Clearly a dimensional reduction $2\mathrm d\to1\mathrm d$ took place\footnote{When $B\to 0$ the typical cyclotron radius $r_l=\sqrt{2l/(eB)}$ becomes infinite, i.e., one has moved to the boundary of the plane.}. Indeed, coming back to  LLL anyons, the claim that in the $B\to 0$ limit the {\rm LLL} anyon thermodynamic becomes  the Calogero thermodynamic is in fact encoded in (\ref{stunning}) since the only input which differentiates them is the 1-body density of states of the said quantum models, the  actual non trivial statistics due  to the anyonic Aharonov-Bohm interactions  in the LLL on the one hand and to the Calogero $1/x^2$ coupling constant on the other hand being encoded \cite{moiter}, \cite{Ouvry1999}, \cite{moibis} in the same common  transcendental equation.

\section*{3) Riemann Zeta function}
As a remark let's focus again on (\ref{LLLbis}) where one has ended up claiming   that
\begin{equation}
\label{LLLter}
Z_{\rm LLL}=\sum_{l=0}^{\infty}\rme^{-\beta\wc}=(1+1+1+\ldots)\;\rme^{-\beta\wc}=\frac{BS}{\phi_0} \rme^{-\beta\wc}
\end{equation}
One can interpret (\ref{LLLter}) by saying that quantum mechanics in the LLL  has given a meaning  to the counting  $1+1+1+\ldots$,   thanks to the harmonic well regularization and  the ensuing thermodynamic limit prescription, by unveiling that  the regularized  $Z_{{\rm LLL},\omega}$ diverges in the thermodynamic  limit $\omega\to 0$ as the surface $S$ of the plane. Now the degenerate sum in  (\ref{LLLter}), up to the factor $\rme^{-\beta\wc}$ that we will from now on ignore as well as the $l=0$ summand, is  nothing but the Riemann Zeta function at zero entry
\begin{equation}\nonumber
\sum_{l=1}^{\infty}\frac{1}{l^0}=1+1+1+\ldots\end{equation}   
 Indeed pushing  to the next order in $\omega$ the  expansion of $Z_{{\rm LLL},\omega}$,  accordingly redefined as a sum without the energy shift $\omega_t$ and starting at  $l=1$, one gets
\begin{equation}\label{omegabis}Z_{{\rm LLL},\omega}=\sum_{l=1}^{\infty}\frac{1}{l^0} \rme^{-\beta l(\omega_t-\omega_c)}
=\frac{2\wc}{\beta\,\omega^2}-\frac{1}{2}+\frac{1}{2\beta\wc}+\mathcal O(\omega^2)
\end{equation}
where on the one hand and as already discussed the leading term ${2\wc/(\beta\,\omega^2)}\to BS/\phi_0$ yields the Landau degeneracy and on the other hand the  zeroth order term   $-{1}/{2}+ {1}/({2\beta\wc})$ is, not surprisingly,  the analytically continued value  \bea\zeta(0)=-1/2\nonumber\eea here shifted by   ${1/(2\beta\wc)}$. So (\ref{omegabis}) contains at least  two informations of interest: on the one hand the actual diverging counting $\sum_{l=1}^{\infty}\frac{1}{l^0}=1+1+1+\ldots$ which  scales\footnote{In the LLL  the degenerate  sum  in (\ref{LLLter}) precisely allows for the degeneracy   $BS/\phi_0$    to factorize from a trivial temperature dependence (equivalently  in (\ref{22})). Note that in the $B\to 0$ limit this would not be the case anymore since  (\ref{88}) (equivalently (\ref{stunning})) clearly leads in the thermodynamic limit to  a spectrum  not degenerate anymore  but quadratic in the 1d quantum number $k$. Pushing the small $\omega$ expansion of $\lim_{B\to0}Z_{{\rm LLL},\omega}$ to the next order in $\omega$ one would get
\begin{equation}\lim_{B\to0}Z_{{\rm LLL},\omega}=\sum_{l=1}^\infty \rme^{-\beta l \omega}
=\frac{1}{\beta\omega}-\frac{1}{2}+\mathcal O(\omega)\nonumber
\end{equation} where again not surprisingly the analytically continued value $\zeta(0)$   materializes as it should but the diverging  ${1}/{\beta\omega}\to {L}/\sqrt{2\pi\beta}$ term  cannot be interpreted as  counting anything anymore. This type of Zeta function regularization  has been extensively used in  the context of the Casimir effect \cite{Casimir}.} as the infinite   surface of the plane and on the other hand its analytically continued value $\zeta(0)=-1/2$.

Riemann Zeta functions regularizations   of this type as well as in other settings (see e.g., \cite{Tao})  are  well-known.  What in addition we learn here  is that the actual divergence in $\sum_{l=1}^{\infty}{1}/{l^0}$ can be interpreted in a  LLL  quantum mechanical context  as a 2 dimensional  space dimension, thanks to a proper thermodynamic limit prescription. 

Following this line of reasoning we define now  for non positive  entries $n\le 0$  the generalized harmonic LLL partition functions
\bea
\label{LLLquar}
Z(n)&=&\sum_{l=1}^{\infty} \frac{1}{l^n}\rme^{-\beta l(\omega_t-\omega_c)}
\eea
as yet another quantum mechanics\footnote{For   example for $n$ even non vanishing   one can rewrite the polylogarithm  in (\ref{LLLquar})     as
  \begin{equation}\nonumber
Z(n)=2^n \frac{\sum_{k=0, k\;{\text {even}}}^{-n-1}
    A(
      \frac{k}{2}, -n) \cosh\big(\frac
       {(n + 1 + k)\beta (\omega_t - \omega_c)}{2}\big)}{\sinh\big(
 \frac{\beta (\omega_t - \omega_c)}{2}\big)^{-n + 1}}
\end{equation}   
   and for $n$ odd   as
\begin{equation}\nonumber
Z(n)=2^n \frac{\sum_{k=0, k\;{\text {even}}}^{-n-3}
    A(
      \frac{k}{2}, -n) \cosh\big(\frac
       {(n + 1 + k)\beta (\omega_t - \omega_c)}{2}\big) + 
   \frac{A\big(\frac{-n - 1}{2}, -n\big)}{2}}{\sinh\big(
 \frac{\beta (\omega_t - \omega_c)}{2}\big)^{-n + 1}}
\end{equation} 
where the  $A(k,n)$'s are the Eulerian numbers
\begin{equation}\nonumber
A(k,n)=\sum _{j=0}^k (-1)^j \binom{n+1}{j} (-j+k+1)^n
\end{equation}} inspired way to regularize   $\sum_{l=1}^{\infty} {1}/{l^n} $  by taking in  the harmonic LLL the $\omega\to 0$ limit along the lines discussed above. 

Now  in quantum mechanics  the factors $l^{-n}$ in (\ref{LLLquar})   can  be interpreted as  the building blocks  of  quantum spectrum degeneracies stemming from adding (similarly to what is done  to obtain the  2d, 3d, $\ldots$ harmonic well spectra from the 1d harmonic well spectrum by adding it 2 times, 3 times, $\ldots$, i.e., one spectrum for each dimension since the Hamiltonians in each space dimension add up)  $-n+1$ times the 1-body harmonic LLL  spectrum $l\;(\omega_t-\omega_c)$, $l\ge 1$, in  which case  the  relevant partition function  ends up being   $Z(0)^{-n+1}$ instead of $Z(n)$. For example $(l_1+l_2)\,(\omega_t-\omega_c)$, $l_1,l_2\ge 1$ is identical to $l\;(\omega_t-\omega_c)$, $l\ge 2$  with degeneracy $l-1$, $(l_1+l_2+l_3)\;(\omega_t-\omega_c)$, $l_1,l_2,l_3\ge 1$ is identical to $l\;(\omega_t-\omega_c)$, $l\ge 3$  with degeneracy $(l - 2) (l - 1)/2$,  $(l_1+l_2+l_3+l_4)\;(\omega_t-\omega_c)$, $l_1,l_2,l_3,l_4\ge 1$ is identical to $l\;(\omega_t-\omega_c)$, $l\ge 4$ with degeneracy $(l - 3) (l - 2) (l - 1)/6$, etc. These degeneracies in turn materialize in the well-known identities  \[Z(0)^2 = Z(-1) - Z(0), Z(0)^3 =\frac{Z(-2) - 3 Z(-1) +  2 Z(0)}{2}, Z(0)^4 = \frac{Z(-3) - 6 Z(-2) + 
    11 Z(-1) - 6 Z(0)}{6}\] etc, with at  zeroth order the combinations\footnote{See OEIS A002208/A002209, a well-known sequence  defined via the integral $\int_{0}^{1}\prod_{k=n}^{0}(k-x)/(1-n)!\;dx$.}  $\zeta(-1)-\zeta(0)=5/12, \big(\zeta(-2) - 3 \zeta(-1) +  2 \zeta(0)\big)/2=-3/8, \big(\zeta(-3) - 6 \zeta(-2) + 
    11 \zeta(-1) - 6 \zeta(0)\big)/6 =251/720, \ldots$ Even though  actual quantum  Landau like Hamiltonians in higher dimensions yielding these spectra are missing,  we still expect that  in the same way  as the  Landau Hamiltonian leads via $Z(0)$ to  a Landau degeneracy 2d surface scaling,   $Z(-1)$ should  lead to a   4d volume scaling, $Z(-2)$ to  a 6d volume scaling, etc, endowing the corresponding Zeta functions with a quantum mechanical space dimension interpretation, in addition to their analytical continued values.

Indeed in view of  the  $\omega\to 0$ expansion $Z(n)$  can  easily be rewritten\footnote{For the sake of completeness we give  the complete $\omega$ expansion \bea\label{hereitis}
Z(n)&=& \zeta(n) + (-1)^n \frac{(-n)!}{ (\beta 2 \omega_c)^{1-n }} + \frac{(1-n )!}{ (\beta 2 \omega_c)^{1-n }} \sum_{k=1}^{1-n}\big(\frac{2 \omega_c}{\omega}\big)^{2 k}
     \frac{\binom{k}{ 1-n - k }}{k}\\\nonumber &+&
       \sum_{k=1}^{\infty}(-1)^k    \big(\frac{\omega}{  2 \omega_c}\big)^{
   2 k} \bigg(\sum_{j=1}^{k}
      \zeta(n - j)  \frac{\binom{2 k - j - 1}{ k - j}}{  
       k (j - 1)! } (\beta 2 \omega_c)^j + (-1)^n \frac{(1-n )!}{  (\beta 2 \omega_c)^{1-n }}
       \frac{\binom{-n + 2 k}{ k}}{  1-n  + k}\bigg)
\eea
where at zeroth order  $\zeta(n)$ is  shifted by  $(-1)^n {(-n)!}/{ (\beta 2 \omega_c)^{1-n }}$.} as
\bea Z(n)&=&(-n)! \frac{1}{\big(\beta(\omega_t-\omega_c)\big)^{1-n}}+\sum_{k= 0}^{\infty}(-1)^k \frac{\big(\beta(\omega_t-\omega_c)\big)^k}{k!}\zeta(n - k)\nonumber\eea where the first term diverges  and the analytically continued value $\zeta(n)$    materializes as expected at zeroth order.

Interestingly enough the most  diverging term tells us, using again the thermodynamic limit prescription (\ref{thecrux}), that in the $\omega\to 0$ limit  $Z(n)$ scales like \bea  (-n)! \frac{1 }{\big(\beta\frac{\omega^2}{2\omega_c}\big)^{1-n}}
     \to  {(-n)!}\big(\frac{ BL^2}{\phi_0}\big)^{1-n}\nonumber\eea i.e., as expected, a $d=2(1-n)$ space dimension stemming from the fact that, as discussed above, one has added $1-n$ times the 2d  LLL spectrum. It follows that a generalization of the  LLL quantum mechanics  has provided interpretation to the divergence in  $\sum_{l=1}^{\infty} {1}/{l^n}$  in terms of  volumes in higher spatial dimensions. It would  certainly be interesting to push further the analysis by giving as well  a meaning, if any,  to the subleading diverging terms in (\ref{hereitis}).

\section*{4) Conclusions}

We have presented a robust thermodynamic limit prescription based on regularizing quantum many-body systems with a smooth harmonic well regulator $\omega \to 0$ rather than traditional  boundary boxes $L \to \infty$. By matching the cluster expansion coefficients of both systems, we derived a precise thermodynamic limit prescription that connects the vanishing harmonic well frequency to an effective infinite spatial volume.
This framework has allowed us to provide a purely quantum-mechanical derivation of the  surface scaling for the lowest Landau level degeneracy without invoking traditional semi-classical or geometric orbital arguments.
By modifying the LLL partition function to include generalized degeneracies, we demonstrated that the mathematical divergences of the Riemann Zeta function $\sum_{l=1}^{\infty} {1}/{l^n}$ at non positive  integer entries  carry direct physical interpretations. Specifically, the leading divergent terms scale  as higher spatial volumes $d = 2(-n+1)$,  mapping number-theoretic regularizations to physical dimensions in quantum mechanics.

Future work will focus on understanding the physical meaning of the subleading divergent terms in the omega expansion, as well as unveiling, if any,  explicit quantum Hamiltonians in higher spatial dimensions that intrinsically mirror these generalized spectra.

\section*{Acknowledgements} The author acknowledges  useful discussions with Olivier Giraud and would like to thank Li Gan for his careful reading of the manuscript.

\end{document}